\documentclass[11pt, a4paper]{article}
\usepackage[T1]{fontenc}
\usepackage[ansinew]{inputenc}
\usepackage{a4wide, url}

\usepackage{authblk}

\usepackage{graphicx}
\usepackage{cite}
\usepackage{amsmath}

\usepackage[normalem]{ulem}
\usepackage{longtable}
\newcommand\suppress[1]{}

\newlength\wvtextpercent
\def\be{\begin{equation}}
\def\ee{\end{equation}}
\def\bea{\begin{eqnarray}}
\def\eea{\end{eqnarray}}

\begin{document}

\title{Thermal lattice Boltzmann method for multiphase flows}

\author[1,2]{Alexander L. Kupershtokh}
\author[1,2]{Dmitry A. Medvedev\footnote{dmedv@hydro.nsc.ru}}
\author[1,2]{Igor I. Gribanov}
\affil[1]{Lavrentyev Institute of Hydrodynamics, Siberian Branch of
Russian Academy of Sciences, Lavrentyev prosp. 15, 630090, Novosibirsk, Russia}
\affil[2]{Novosibirsk State University, Pirogova str. 2, 630090, Novosibirsk, Russia}

\date{\today}

\maketitle

\begin{abstract}
New method to simulate heat transport in the multiphase lattice Boltzmann (LB) method is
proposed. The energy transport equation needs to be solved when phase boundaries are
present. Internal energy is represented by an additional set of distribution functions, which
evolve according to a LB-like equation simulating the transport of a passive scalar. Parasitic heat
diffusion near boundaries with large density gradient is supressed by using the
interparticle ``pseudoforces'' which prevent the spreading of energy. The compression work and heat
diffusion are calculated by finite differences. The latent heat of a phase transition is released
or absorbed inside of a thin transition layer between liquid and vapor. This
allows one to avoide the interface tracking. Several tests are carried out concerning all aspects
of the processes. It is shown that the Galilean invariance and the scaling of thermal conduction
process hold as well as the correct dependence of sound speed on the heat capacity ratio. The
method proposed has low scheme diffusion of the internal energy, and it can be applied to modeling
a wide range of multiphase flows with heat and mass transfer.
\end{abstract}

\section{Introduction}\label{intro}

Simulation of fluid flows with phase transitions between liquid and vapor is difficult because new
phase boundaries can apper in the bulk during calculations, and the existing boundaries can
disappear or change their topology. Therefore, the application of interface tracking methods is
difficult if not impossible. Moreover, the density ratio of liquid and vapor phases is usually high
(can reach tens and hundreds of thousands) leading to noticeable numerical diffusion and/or
dispesion near the boundaries when using common finite-difference methods.

The lattice Boltzmann method (LBM) \cite{McNamZan88,HigJim89} is based on the solution of a
kinetic equation for pseudoparticles. It was widely applied for simulating flows of single-phase
and multiphase media \cite{ChenDoolen98,aidun10,ShanChen93,QianChen97,PRE2006Kup,kuper12,kuper14}.
Moreover, the method is easily parallelizable on graphic accelerators using the GUDA technology
\cite{kuper12,kuper14,LiWei03,tolke08,obrecht11,walsh13}.

Three essentially different approaches were proposed to simulate the heat transport in the LBM: the
model with the extended velocity set \cite{AlexChenSterl93,Qian93,ChenOhaAki94,RendaEtal98}, the use of
the  finite-difference equation for energy \cite{zhang03}, and the method of a second set of
distribution functions (method of passive scalar) \cite{Shan97,HeChenDoolen98}.

The first approach has rather narrow range of simulated temperature where simulations are stable.
Moreover, the size of data increases significantly. This method was applied only to single-phase flows.
When the energy equation is solved by finite difference methods, the large numeric diffusion arises in
the case of moving fluid that restricts significantly the possibilty of modeling.

The passive scalar approach is realized in the LBM by introducing the additional set of
distribution functions. This model has much lower numeric diffusion than the finite-difference
method. Usually the temperature $T$ is used as a passive scalar which can be done only in a case of
almost constant density of the fluid. When a phase transition liquid-vapor is present, the change
of density is not small, and the tarnsport of internal energy should be considered instead of
temperature. Such approach was used in the works \cite{HeChenDoolen98,guo07}, and it was combined
with extended velocity set in the work \cite{li07}. However, only single-phase flows were
considered in these works.

\section{Lattice Boltzmann method}\label{s:method}

The lattice Boltzmann method (LBM) is based on the solution of a kinetic equation for
pseudoparticles. Only a limited set of particle velocities ${\bf c}_k$ is possible such that the
vectors ${\bf e}_k={\bf c}_k\Delta t$ correspond to vectors to neighbor nodes of a regular spatial
lattice \cite{QianDHumLall92}. The usual choice of velocity set is three vectors in one-dimensional
case $|{\bf c}_k|=0, h/\Delta t$, D1Q3 model, nine velocities in 2D case (D2Q9, $|{\bf c}_k|=0,
h/\Delta t,\sqrt{2}h/\Delta t$), and nineteen velocities in 3D case (D3Q19, $|{\bf c}_k|=0,
h/\Delta t,\sqrt{2}h/\Delta t$). Here, $h$ is the lattice space, $\Delta t$ is time step.

One-particle distribution functions $N_k$ are used as main variables, they have the meaning of
the parts of fluid density. Evolution equation for $N_k$ has the form
\begin{equation}
 N_k({\bf x}+{\bf c}_k\Delta t,t+\Delta t)=N_k({\bf x},t)+\Omega_k(N)+\Delta N_k,\quad k=0,\ldots,b,
\label{eq:lbe}
\end{equation}
where $\Omega_k$ is the collision operator, and $\Delta N_k$ is the change of distribution
functions under the action of volume forces (both external and internal).

The collision operator is mostly chosen in the form of a relaxation to local equilibrium with one
(BGK, Bhatnagar-Gross-Krook model \cite{QianDHumLall92}) or several relaxation times (MRT,
Multi-Relaxation-Time model \cite{lall00}). For the BGK model, the collision operator is
$$
 \Omega_k=\frac{N_k^{eq}(\rho,{\bf u})-N_k({\bf x},t)}{\tau},
$$
where $\tau=t_{rel}/\Delta t$ is the non-dimensional relaxation time.
Equilibrium distribution functions are usually taken as truncated Maxwellians up to the second
order in fluid velocity $u$ \cite{koelm91}
\begin{equation}
 N_k^{eq}(\rho,{\bf u})=\rho w_k\left(1+\frac{({\bf c}_k{\bf u})}{\theta}+\frac{({\bf c}_k{\bf
u})^2}{2\theta^2}-\frac{u^2}{2\theta}\right).\label{eq:feq}
\end{equation}
Coefficients $w_k$ depend on the lattice geometry \cite{QianDHumLall92}. For the one-dimensional
model D1Q3 they are $w_0=2/3, w_{1-2}=1/6$, for two-dimensional model D2Q9 $w_0=4/9, w_{1-4}=1/9,
w_{5-9}=1/36$, and for three-dimensional model D3Q19 $w_0=1/3, w_{21-6}=1/18, w_{7-18}=1/36$. The
kinetic temperature of pseudoparticles in LB models listed is $\theta=(h/\Delta t)^2/3$, and the
kinematic viscosity is defined by the relaxation time $\nu=\theta(\tau-1/2)/\Delta t$.
Change of distribution functions at a node is calculated using the Exact Difference Method (EDM)
\cite{kuper04b}
\begin{equation}
 \Delta N_k({\bf x}, t)=N_k^{eq}(\rho,{\bf u}+\Delta{\bf u})-N_k^{eq}(\rho,{\bf u})\label{eq:edm},
\end{equation}
where $\Delta{\bf u}={\bf F}\Delta t/\rho$ is the change of fluid velocity in one time step, and
{\bf F} is the total force acting on the fluid at a node.

Density $\rho$ and velocity {\bf u} of fluid are calculated as
\begin{equation}
 \rho=\sum\limits_{k=0}^b N_k,\quad \rho{\bf u}=\sum\limits_{k=1}^b {\bf c}_kN_k.\label{eq:hydro}
\end{equation}

Under the action of volume forces, the physical fluid velocity ${\bf u}^*$ is defined at half time
step \cite{ginzb94}
\begin{equation}
 \rho{\bf u}^*=\sum\limits_{k=1}^b{\bf c}_kN_k+{\bf F}\Delta t/2.\label{eq:ustar}
\end{equation}

\section{Phase transitions}

In order to obtain phase transitions in a fluid, it is necessary to model in the LBM the attractive
part of the ``intermolecular'' interaction. This was done in the work \cite{ShanChen93} by the
introduction of attractive forces acting on the fluid at a node from neighbor nodes. Later, the
total force {\bf F} acting on the fluid at a node was introduced as a gradient of pseudopotential
$U$ defined from the equation of state for the fluid \cite{QianChen97,zhang03}
\begin{equation}
 {\bf F}=-\nabla U=-\nabla(p(\rho,T)-\rho\theta).\label{eq:force}
\end{equation}
LBM with such attractive forces represents phase boundary as a thin transition layer between liquid
and vapor where density changes smoothly across several lattice nodes (interface capturing). The
surface tension arises at phase boundaries.

In the work \cite{kuper05e}, a new function was introduced $\Phi=\sqrt{-U}$. The equation
(\ref{eq:force}) can be rewritten in the equivalent form
\begin{equation}
 {\bf F}_N=2A\nabla(\Phi^2)+(1-2A)2\Phi\nabla\Phi\label{eq:force2},
\end{equation}
where $A$ is a free parameter which allows one to obtain correct phase densities at the coexistence
curve. The sufficiently isotropic approximation of the formula (\ref{eq:force2}) is
$$
{\bf F}({\bf x})=\frac{1}{\alpha h}\left[A\sum\limits_{k-1}^b G_k\Phi^2({\bf x}+{\bf e}_k){\bf
e}_k+(1-2A)\Phi({\bf x})\sum\limits_{k-1}^b G_k\Phi({\bf x}+{\bf e}_k){\bf
e}_k\right],$$
where coefficients $G_k>0$ differ for different lattice directions. For neighbor nodes, they are
$G_k=1$. For next-neighbor nodes, the vaues of coefficient ensuring isotropy are $G_{5-8}=1/4$ in
the two-dimensional model D2Q9 and $G_{7-18}=1/2$ in the three-dimensional model D3Q19. The
coefficient $\alpha$ is equal to 1, $3/2$, and 3 for the models D1Q3, D2Q9, and D3Q19,
correspondingly.

In present work, we used Van der Waals equation of state which is written in reduced variables as
$$
\tilde{p}=\frac{8\tilde{\rho}\tilde{T}}{3-\tilde{\rho}}-3\tilde{\rho}^2.
$$
Here and below, pressure, density and temperature are scaled by their values at the critical point,
$\tilde{p}=p/p_c, \tilde{\rho}=\rho/\rho_c, \tilde{T}=T/T_c$. For this equation, the approximation
(\ref{eq:force2}) gives the best agreement with the phase coexistence curve at $A=-0.152$ (the
deviation of density in simulations from the theoretical value is less than 0.2\% in the range of
temperature from the critical one $\tilde{T}=1$ down to $\tilde{T}=0.4$ \cite{kuper05e}). More
complex equations of state including tabulated ones for real fluids were considered in
\cite{cma2009,kuper11}.

The stability of the LBM with equation of state in the form $p=p(\rho,T)$ is defined by the
criterion \cite{kuper10}
$$
\left(\frac{\partial p}{\partial\rho}\right)_T\leq\left(\frac{h}{\Delta t}\right)^2+\theta.
$$

\section{Heat transport}\label{s:heat}

Evolution equation for the internal energy per unit volume $E$ is
\begin{equation}
  \frac{\partial E}{\partial t}+\nabla\cdot({\bf u}E)=\frac{p}{\rho}\frac{d\rho}{dt} +
\nabla\cdot(\lambda\nabla T) + \hat{\sigma}:\nabla{\bf u},\label{eq:energy}
\end{equation}
where the first term in the right hand side corresponds to the pressure work, the second one represents the
heat conduction, and the last one is the viscous heating. Here, $\lambda=\rho C_V\chi$ is the heat
conductivity, $C_V$ is the specific heat at constant volume, $\chi$ is the thermal diffusivity,
and $\hat{\sigma}$ is the viscous stress tensor. It is more convenient to express the pressure work
from the velocity divergence using the
continuity equation
$$
  \frac{p}{\rho}\frac{d\rho}{dt}=-p\operatorname{div}({\bf u}^*).
$$
Calculation of the left hand side of Eq.~(\ref{eq:energy}) is the most complicated, i.e., the advection of
the internal energy by the fluid flow with velocity calculated from Eq.~(\ref{eq:ustar}). The
viscous heating is usually small, and we will neglect it in the following.

This was done in the works \cite{AlexChenSterl93,Qian93,ChenOhaAki94,RendaEtal98} by introducing
the extended set of velocity vectors ${\bf c}_k$ and the increasing the expension order of
euilibrium distribution functions (terms up to fourth order in {\bf u} were used). The drawbacks of
this approach is relatively narrow range of simulated temperature where simulations are stable,
and the significant increase of the size of data.

In the work \cite{zhang03}, the advection of energy in equation (\ref{eq:energy}) was calculated by
a finite-difference method using the values of fluid density and velocity obtained from LBE.
However, this method produces large numerical diffusion and dispersion of energy near phase
boundaries in simulations of moving fluids which significantly complicates modeling.

Third approach to simulating the advection of energy in LBM is based on the use of passive scalar
(additional set of distribution functions $g_k$) \cite{Shan97} which has much lower scheme diffusion
comparing to finite-difference methods. Earlier, this approch was used for simulating the flows
with almost constant fluid density and specific heat when temperature can be used as passive
scalar. At phase transitions between liquid and vapor, the change of density is however not small,
and the advection of internal energy should be considered instead of temperature.

When using the passive scalar approach $E=\sum\limits_{k=0}^b g_k$ for the internal energy density
$E=\rho c_VT$, evolution equations for distribution functions $g_k$ can be written in the form
analogous to Eq.~(\ref{eq:lbe})
\begin{equation}
 g_k({\bf x}+{\bf c}_k\Delta t,t+\Delta t)=g_k({\bf x},t)+\frac{g_k^{eq}(E,{\bf u})-g_k({\bf
x},t)}{\tau_E}+\Delta g_k({\bf x},t).\label{eq:gk}
\end{equation}
Here, $\Delta g_k=\Delta g_k^{(1)}+\Delta g_k^{(2)}$ is the total change of distribution functions,
$\tau_E$ is the non-dimensional relaxation time for the energy. Equilibrium distribution functions
$g_k^{eq}(E,{\bf u})$ have the same form as $N_k^{eq}(E,{\bf u})$ (\ref{eq:feq}).

The change of energy at a node $\Delta E$ due to the pressure work and the heat conduction is
calculated by usual finite-difference formulas. Corresponding changes of energy distribution
functions $\Delta g_k^{(1)}$ are proportional to the change of energy
$$
\Delta g_k^{(1)}({\bf x},t)=g_k({\bf x},t)\frac{\Delta E}{E}.
$$

The main problem in this approach stems from the jump of specific heat at phase boundaries. This
leads to parasitic diffusion (spreading) of internal energy from dense phase (liquid) to rarefied
one (vapor) even if pressure and temperature are uniform. This effect is readily observed for
stationary droplet in the case of barotropic equation of state (pressure depends only on density).
Since there is no feedback on temperature, waves of pressure and density does not arise. To
demonstrate the parasitic diffusion of energy, the barotropic van der Waals equation of state was
used with a constant $\tilde{T}_0$ instead of the temperature of fluid
$$
\tilde{p}=\frac{8\tilde{\rho}\tilde{T}_0}{3-\tilde{\rho}}-3\tilde{\rho}^2.
$$

\begin{figure}[t]
\noindent
\centering
 \includegraphics[width=\textwidth]{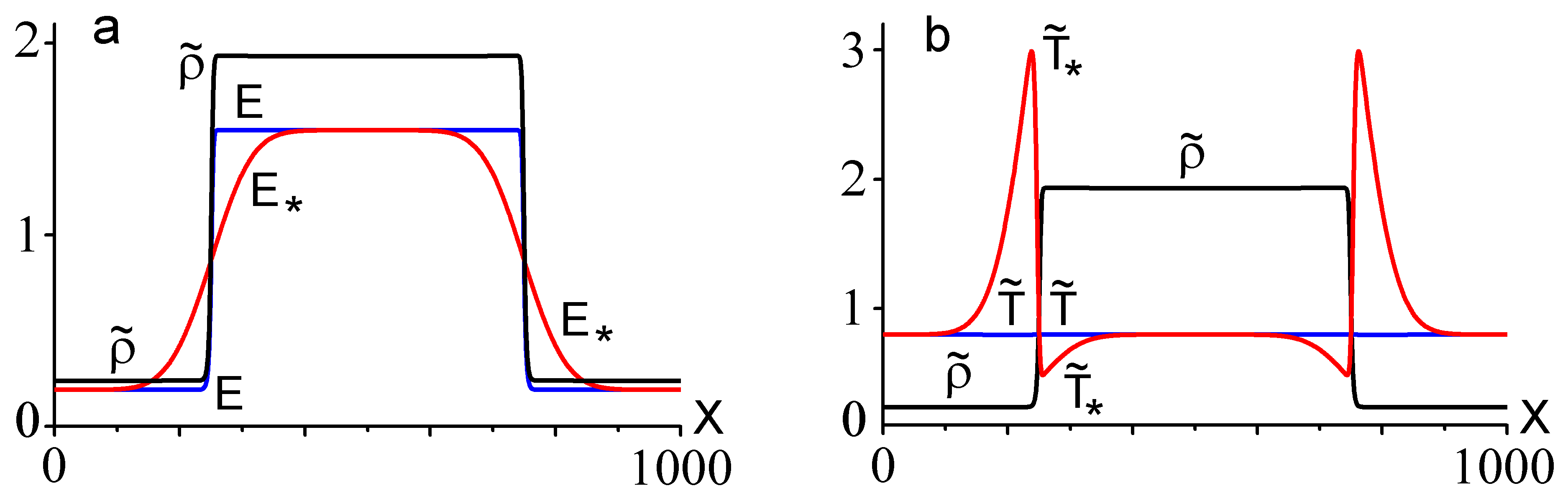}
 \caption{Stationary one-dimensional droplet in saturated vapor. Parasitic diffusion (spreading) of
energy at phase boundaries without use of ``pseudoforces''.\label{f:spread}}
\end{figure}

Parasitic diffusion at phase boundaries is shown in Fig.~\ref{f:spread}. The pressure work and heat
diffusion were switched off for clarity. Internal energy ``leaks'' from the stationary liquid
droplet to the surrounding saturated vapor. This leads to generation of non-physical temperature
peaks in vapor and drops in liquid near the boundaries. In thermal simulations, such peaks and
drops will lead to instability.

\begin{figure}[t]
\noindent
\centering
 \includegraphics[width=8cm]{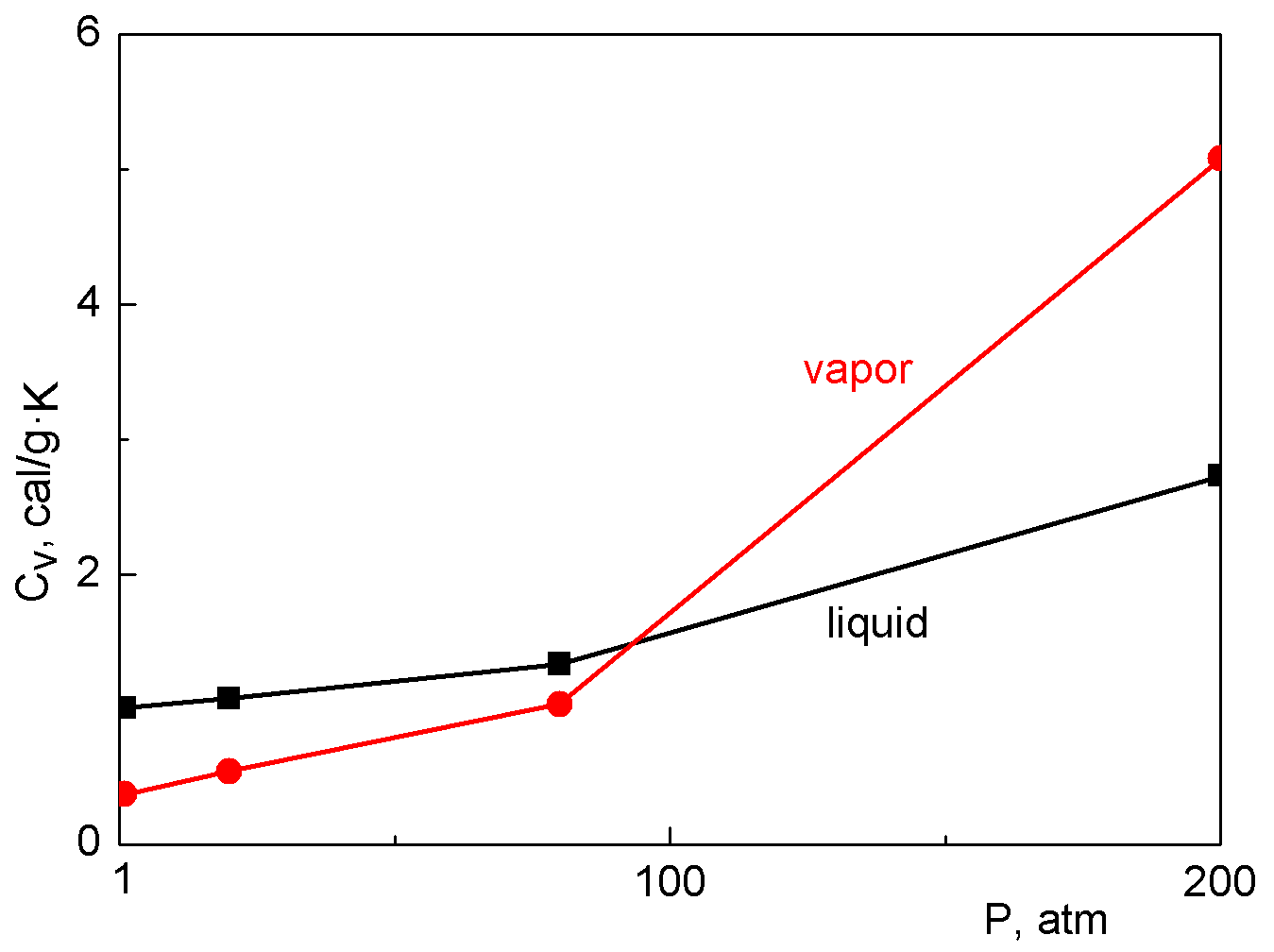}
 \caption{Specific heat of water and vapor near the saturation curve.\label{f:spec-heat}}
\end{figure}

In this work, we modify the passive scalar approach in order to be used for the advection
of internal energy. The idea is to introduce special ``pseudoforces'' for energy scalar which
prevent the spreading at phase boundaries. The currently realized variant works in the case of a
constant specific heat of fluid $C_V$, and the internal energy at given temperature proportional
to the fluid density. This is valid for van der Waals and other linear in temperature equations of
state since for them
$$
\left(\frac{\partial E}{\partial V}\right)_T=P\left(\frac{\partial P}{\partial T}\right)_V-P=0.
$$
For water, the liquid and vapor specific heat along the coexistence curve are also close in a
certain range of pressure (see Fig.~\ref{f:spec-heat}). ``Pseudoforces'' are taken into account in
the evolution equation for distribution functions (\ref{eq:gk}) by the EDM similar to
Eq.~(\ref{eq:edm})
$$
 \Delta g_k^{(2)}({\bf x},t)=g_k^{eq}(E,{\bf u}+\Delta{\bf u})-g_k^{eq}(E,{\bf u}).
$$
Here {\bf u} is the fluid velocity defined by the main set of distribution functions
(\ref{eq:hydro}).

\section{Latent heat of phase transition}\label{s:lheat}

\begin{figure}[t]
 \includegraphics[width=\textwidth]{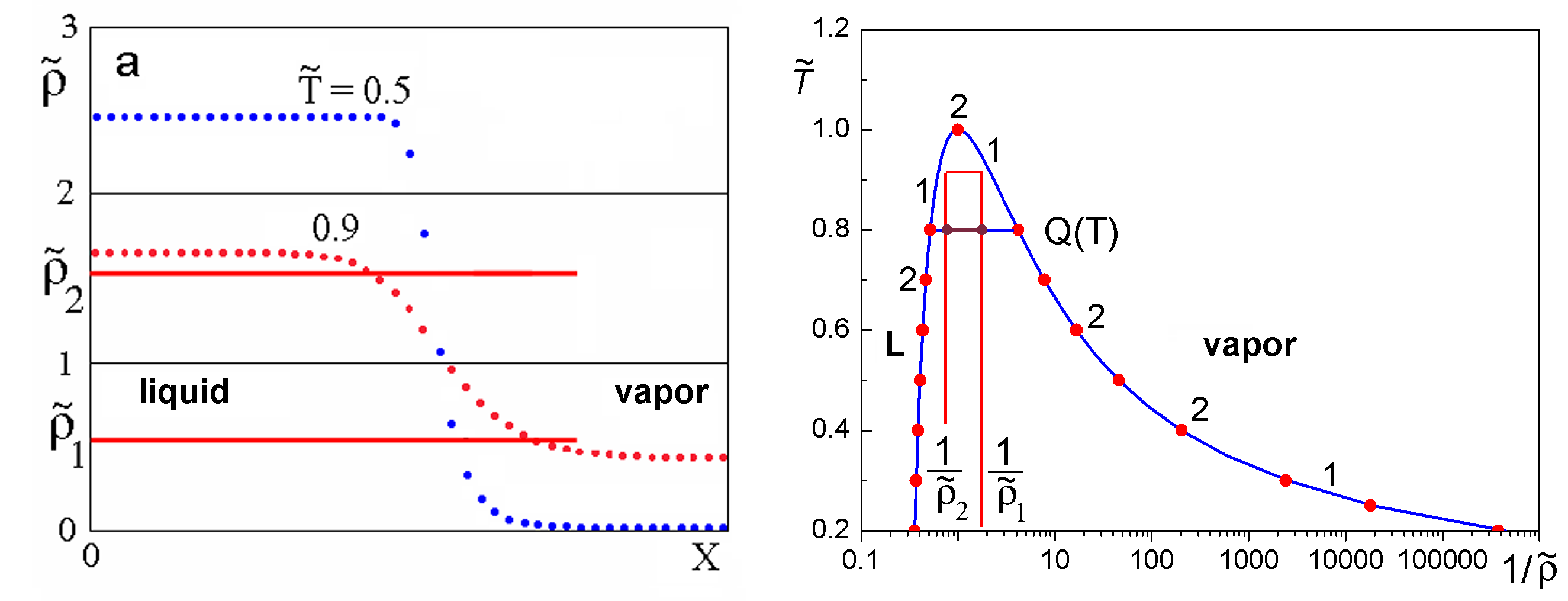}
 \caption{Schematic of taking into account the latent heat of phase transition.\label{f:lhscheme}}
\end{figure}

It is known that the latent heat of phase transition should be taken into account at the
conditions at a moving phase boundary. Corresponding boundary conditions are
$$
 \lambda_{liq}\left.\frac{\partial T}{\partial
x}\right|_{x=\xi-0}-\lambda_{vap}\left.\frac{\partial T}{\partial x}\right|_{x=\xi+0}
=\rho_{liq}(T)Q(T)\frac{d\xi}{dt},
$$
where $\xi$ is the coordinate of a planar pahse boundary between liquid and vapor, and
$\rho_{liq}(T)$ is the liquid density at the phase coexistence curve (see Fig.~\ref{f:lhscheme},{\it
b}). The latent heat of phase transition $Q(T)$ decreases with the increasing temperature, and
comes to zero at the critical temperature $T=T_c$. Tracking the phase boundaries is difficult in
simulations because in many cases new boundaries can appear, and existing ones can disappear or
change their topology. The advantage of the LBM is its capturing of interfaces. Phase boundaries in
the LBM are represented as transition layers where the density continuously change from the liquid
to the vapor values according to the phase coexistence curve. The density of a portion of fluid at
a phase transition also changes continuously in time. We propose to take into account the latent
heat in the following way. If we do not resolve exactly the inner structure of the transition layer
(see Fig.~\ref{f:lhscheme}{\it a}) but take into account the latent heat of phase transition only
integrally accross the transition layer, we can assume that the latent heat is released or absorbed
continuously inside the transition layer in a certain range of density $\rho_1<\rho<\rho_2$ (Fig.~
\ref{f:lhscheme}{\it b}) according to the equation
$$
\frac{dE}{dt}=\frac{\rho_{liq}Q(T)}{\rho_2-\rho_1}\frac{d\rho}{dt}=-\frac{\rho_{liq}Q(T)}{
\rho_2-\rho_1}\rho\operatorname{div}({\bf u}^*).
$$
Equilibrium densities of the vapor and the liquid $\rho_{liq}$ at every temperature can be used as
$\rho_1$ and $\rho_2$, correspondingly.

\section{Numerical validation}\label{numeric}

\subsection{Galilean invariance and scheme diffusion of energy}

\begin{figure}[t]
\noindent
\centering
\includegraphics[width=\textwidth]{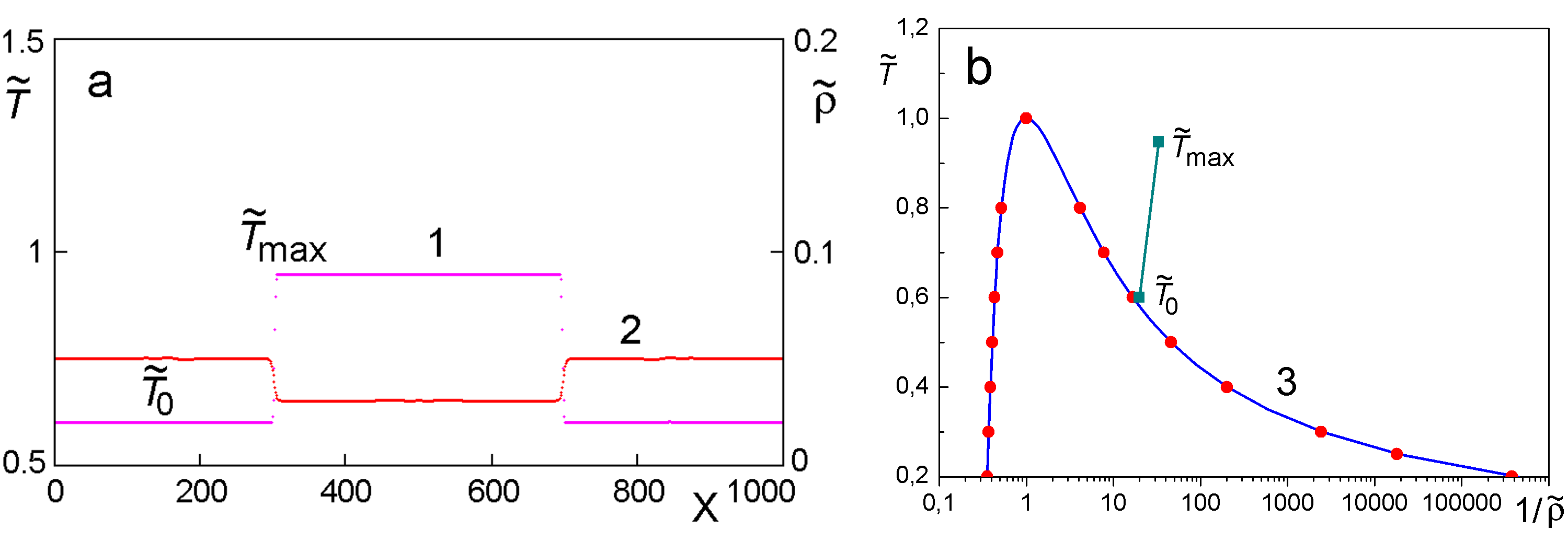}
\caption{Initial state ({\it a}): distribution of temperature (curve 1) and density (curve 2).
Phase coexistence curve ({\it b}): theoretical calculations by Maxwell rule (curve 3) and LBM
simulation results (points). $\tilde{T}_0=0.6, \tilde{\rho}_0=0.05, \tilde{T}_{max}=0.947$
\label{f:init}}
\end{figure}

The initial state for the tests of Galilean invariance and scheme diffusion is shown in
Fig.~\ref{f:init}. The temperature and the density were distributed stepwise, and the pressure was
constant. Periodic boundary conditions were used. The coefficient of scheme diffusion for energy was
$D_E=\theta(\tau_E-1/2)\Delta t=0.001h^2/\Delta t$ with $\tau_E=0.503$.

\begin{figure}[t]
\noindent
\centering
 \includegraphics[width=\textwidth]{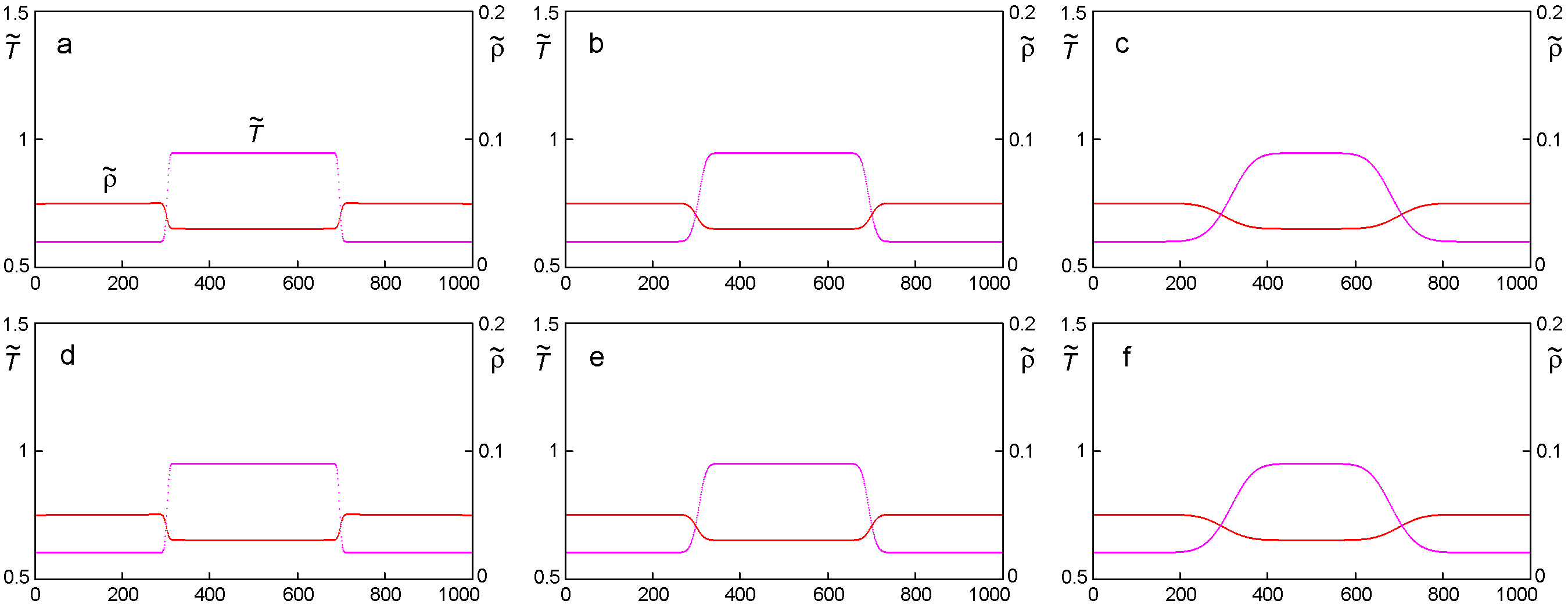}
 \caption{Scheme diffusion of energy for flow velocity $u=0$ ({\it a}--{\it c}) and
$u=0.1h/\Delta t$ ({\it d}--{\it f}). Time $t=10000$ ({\it a}, {\it d}), $100000$
({\it b}, {\it e}), $1000000$ ({\it c}, {\it f}). \label{f:gali}}
\end{figure}

Figure~\ref{f:gali} shows distribution of temperature and density at diffrent time in the case of
zero flow velocity ({\it a}--{\it c}), and for uniform flow velocity equal to $u=0.1h/\Delta t$
({\it d}--{\it f}). Results are independent on the flow velocity, hence, the Galilean invariance
holds.

\begin{figure}[t]
\noindent
\centering
 \includegraphics[width=\textwidth]{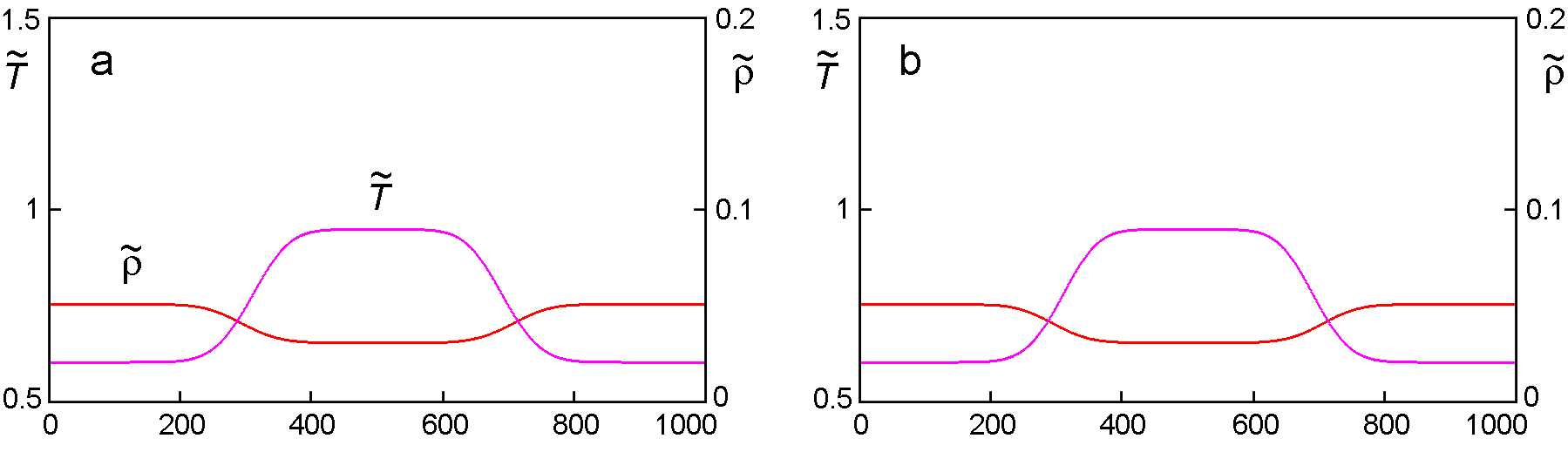}
 \caption{{\it a} -- thermal diffusivity $\chi=0.01h^2/\Delta t$, $t=100000$;
{\it b} -- thermal diffusivity $\chi=0.1h^2/\Delta t$, $t=10000$.\label{f:diffus}}
\end{figure}

Figure~\ref{f:diffus} shows the distribution of temperature in a resting fluid for two different
thermal diffusivity and corresponding time. The similarity relation $l\sim\sqrt{\chi t}$ is
fulfilled.

Since an explicit numerical scheme was used for calculating the heat diffusion, the stability
criterion is $\chi\Delta t/h^2<0.5/d$ where $d$ is the number of spatial dimensions.
One-dimensional calculations with the flow velocity equal to $u=0.1h/\Delta t$ and $\chi\Delta
t/h^2=0.49$ were indeed stable.

Two-dimensional simulations were carried out using the D2Q9 model. A round droplet surrounded by a
saturated vapor moved with uniform velocity along the diagonal of the simulation region. Periodic
boundary conditions were used for both $x$ and $y$ directions. The initial temperature was
constant, hence, the density of internal energy was higher inside the droplet. During the
simulation time of $t=62000$, the droplet made more than six revolutions together with the flow
which corresponds to 27 droplet diameters. The isotropy is preserved (the droplet remained round),
and almost no parasitic diffusion of energy was present. Three-dimensional simulations with D3Q19
model gave similar results.

\begin{figure}[t]
\noindent
\centering
 \includegraphics[width=\textwidth]{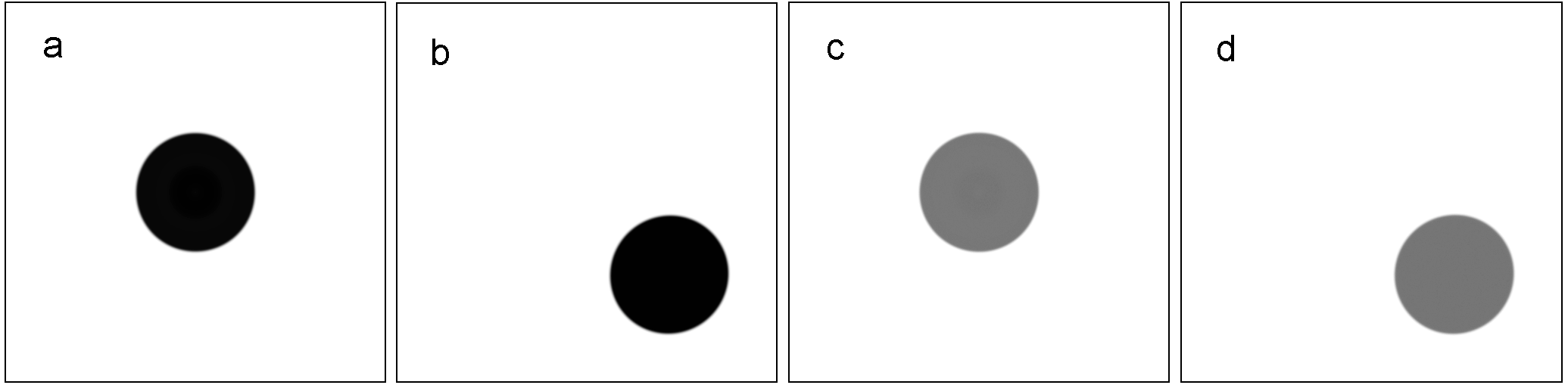}
 \caption{Moving two-dimensional liquid droplet of radius $R=160$ in saturated vapor. Distribution
of density ({\it a}, {\it b}) and internal energy ({\it c}, {\it d}). Flow velocity
$u_x=0.1h/\Delta t, u_y=-0.1h/\Delta t$. Time $t=0$ ({\it a}, {\it c}), $62000$ ({\it b}, {\it
d}). Grid size $1000\times 1000$.\label{f:drop2d}}
\end{figure}

\subsection{Pressure work}

In order to check the calculations of pressure work, we investigated the dependence of the speed of
sound on the specific heat. If pressure work is switched off, one obtains the isothermal speed of
sound $c_T$, and taking into account the pressure work gives the adiabatic speed of sound $c_S$.
For Van der Waals equation of state, the reduced values of both speeds at the temperature
$\tilde{T}_0$ are

\begin{equation}
c_T=\left(\frac{\partial\tilde{p}}{\partial\tilde{\rho}}\right)_T=\frac{24\tilde{T}_0}{
(3-\tilde{\rho})^2}-6\tilde{\rho},\quad
c_S=\left(\frac{\partial\tilde{p}}{\partial\tilde{\rho}}\right)_S=\frac{24\gamma\tilde{T}_0}{
(3-\tilde{\rho})^2}-6\tilde{\rho}. \label{eq:sspeed}
\end{equation}

\begin{figure}[t]
\noindent
\centering
 \includegraphics[width=8cm]{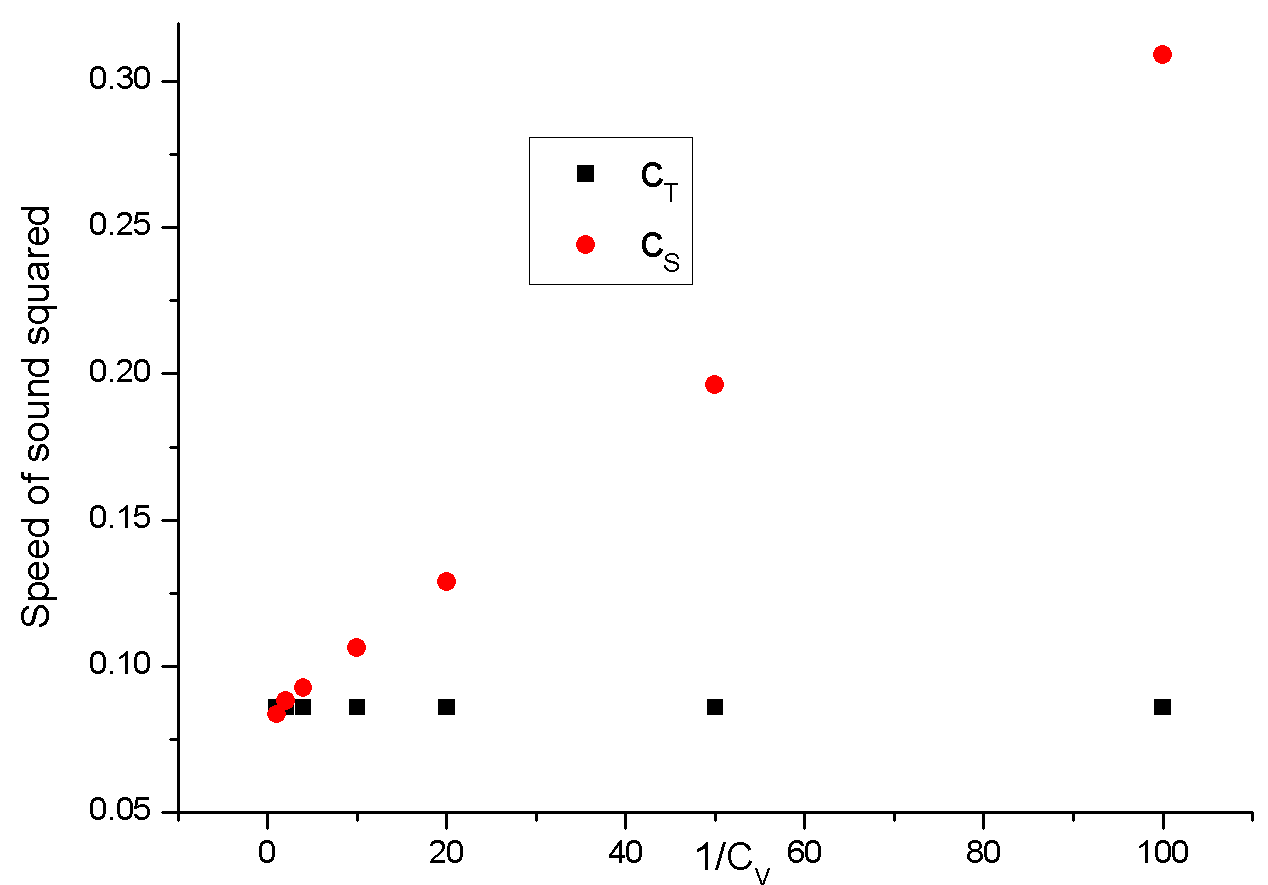}
 \caption{Speed of sound vs. inverse specific heat. Density $\tilde{\rho}=0.01$, temperature
$\tilde{T}=0.8$.\label{f:sspeed}}
\end{figure}

Here, $\gamma=C_P/C_V$ is the heat capacity ratio. The speed of sound was calculated from the
dispersion relation for a standing harmonic wave, $c=\omega L/2\pi$ where $L$ is the wavelength,
and $\omega$ is the frequency. Figure~\ref{f:sspeed} shows the dependence of the speed of sound on
the inverse specific heat $C_V$. The isothermal speed of sound is constant, and the adiabatic speed
of sound depends linearly on $1/C_V$ with agreement of the theoretical result (\ref{eq:sspeed}).

\begin{figure}[t]
\noindent
\centering
 \includegraphics[width=\textwidth]{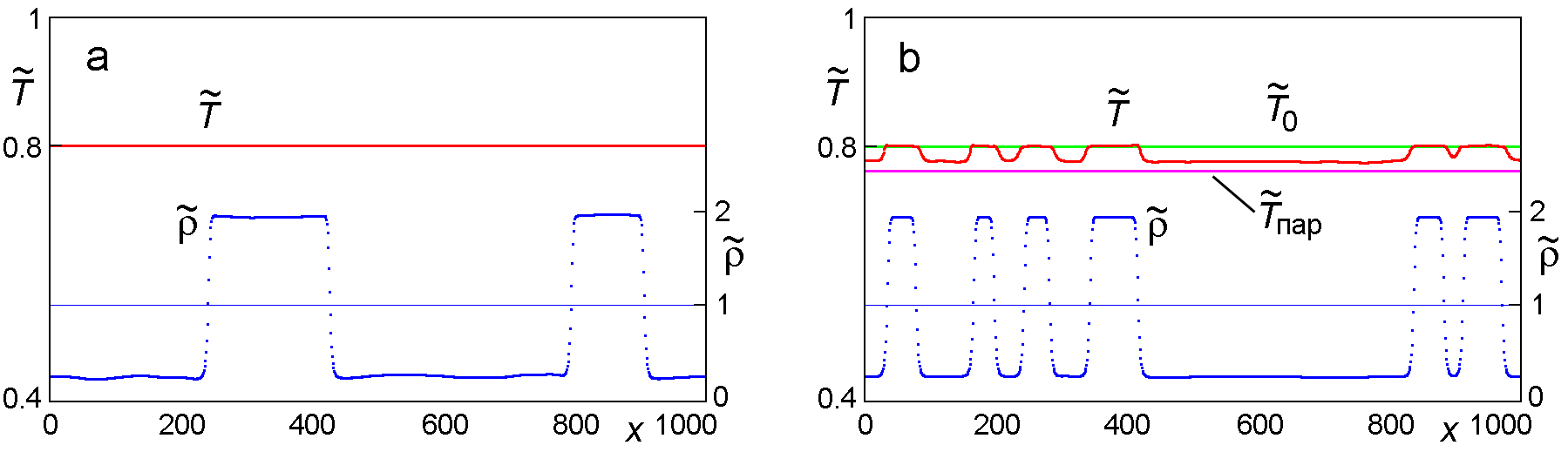}
 \caption{Distribution of the temperature and the density of fluid after the spinodal
decomposition. Pressure work is neglected ({\it a}) and taken into account ({\it
b}). $\tilde{T}_0=0.8, \tilde{\rho}_0=0.7$, $t=50000$.\label{f:spinodal}}
\end{figure}

Another test was the simulation of a spinodal decomposition (decay of an initially uniform
fluid which is in thermodynamic state below the spinodal into a mixture of liquid and vapor).
Figure~\ref{f:spinodal} shows the simulation results for the case of neglected pressure work
(Fig.~\ref{f:spinodal}{\it a}), and the case of pressure work taken into account
(Fig.~\ref{f:spinodal}{\it b}). The pressure work was calculated with a finite-difference expression
$$
\Delta E_i^n=-p_i^n\frac{(u^*)_{i+1}^n-(u^*)_{i-1}^n}{2h}\Delta t.
$$
Thermal diffusivity was set to zero, and only small scheme diffusion was present.
Without the pressure work, temperature remains constant. The pressure work results in the increase
of the internal energy in liquid phase (which is compressed) and decrease of the internal energy in
gas phase (where rarefaction occurs). Since the compression of liquid is relatively low, and the
specific heat is significantly larger than that of the vapor, the temperature of the liquid
increases only slightly. In contrast, the temperature of the gas phase decreases significantly. One
can estimate the change of the vapor temperature as
$$
T_{vap}=T_0-\frac{p}{C_V}\left(\frac{1}{\rho_{vap}}-\frac{1}{\rho_0}\right).
$$
From Fig.~\ref{f:spinodal} one can see that simulations give close value of the vapor temperature
(Fig.~\ref{f:spinodal}{\it b}).

\subsection{Latent heat of phase transition}

\begin{figure}[t]
\noindent
\centering
 \includegraphics[width=\textwidth]{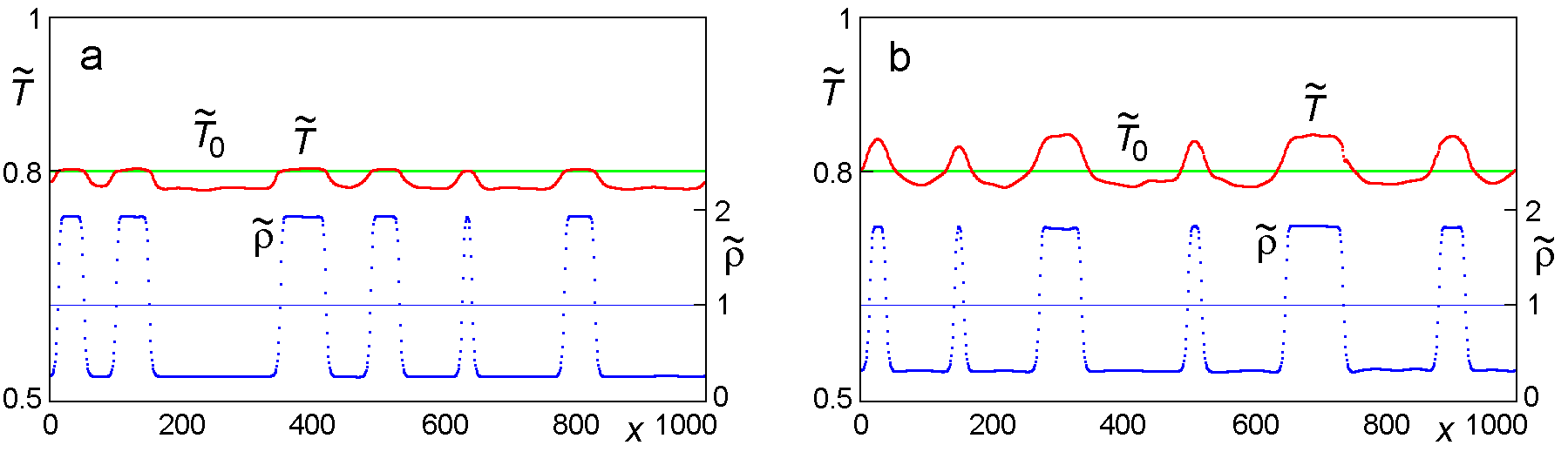}
 \caption{Distribution of the temperature and the density of fluid after the spinodal
decomposition. Latent heat is $\tilde{Q}=0$ ({\it a}) and $\tilde{Q}=0.02$ ({\it b}). Initial
temperature $\tilde{T}_0=0.8, \tilde{\rho}_0=0.7$, $t=50000$.\label{f:lh}}
\end{figure}

The case of spinodal decomposition was simulated with zero and non-zero latent heat of phase
transition. Results are shown in Fig.~\ref{f:lh}. When the latent heat is non-zero, the temperature
of liquid phase is significantly higher than the initial one due to the release of the latent heat
at the condensation of vapor.

\begin{figure}[t]
\noindent
\centering
 \includegraphics[width=0.7\textwidth]{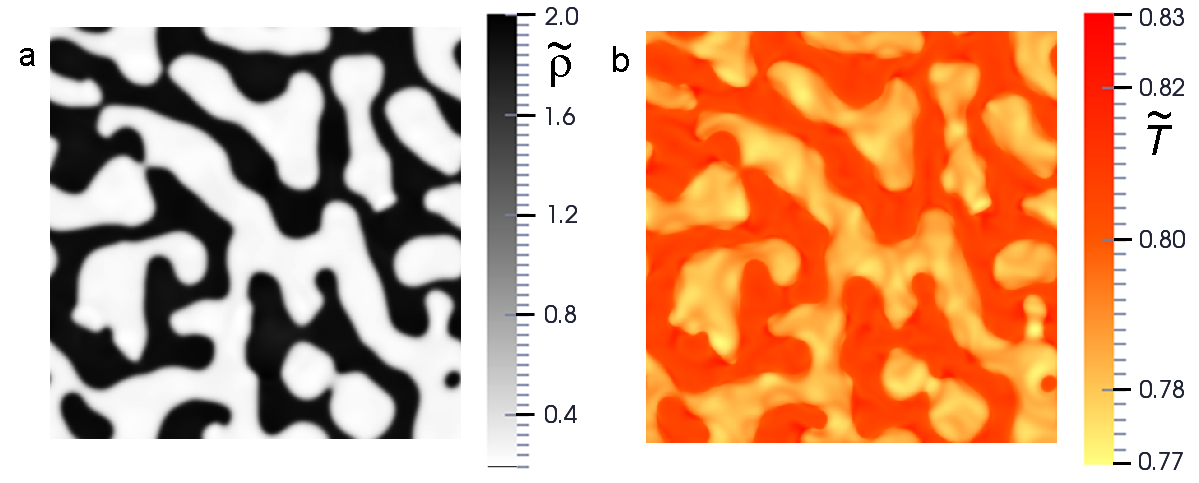}
 \caption{Spinodal decomposition. Distribution of density ({\it a}) and
temperature ({\it b}). $\tilde{T}_0=0.8, \tilde{\rho}_0=1, \tilde{\rho}_{liq}\tilde{Q}=0.02$. Grid
size $500\times 500$, $t=2630$.\label{f:lh2d}}

\vspace*{5mm}
\includegraphics[width=0.7\textwidth]{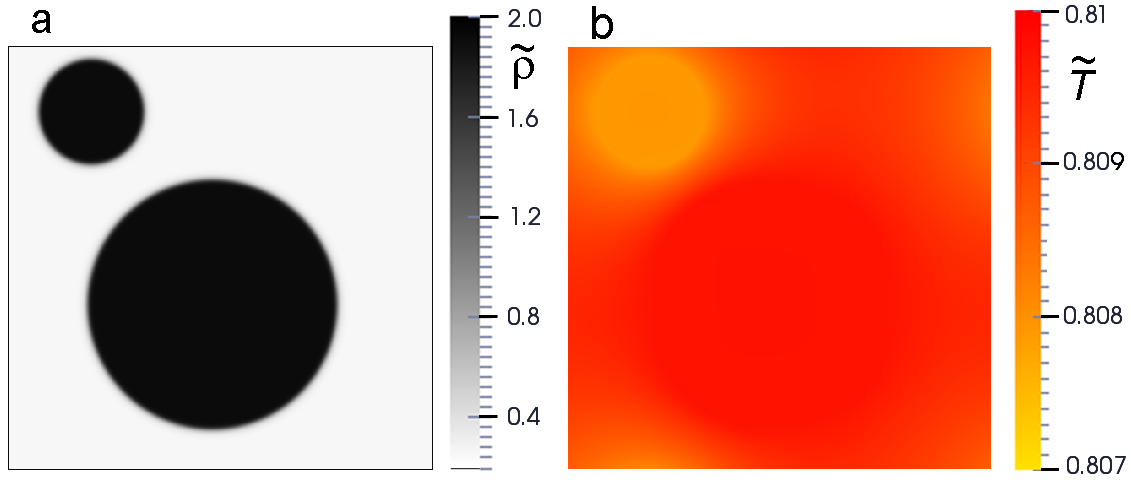}
 \caption{Distribution of density ({\it a}) and temperature ({\it b}) at
the late stage of spinodal decomposition. $\tilde{T}_0=0.8, \tilde{\rho}_0=0.8,
\tilde{\rho}_{liq}\tilde{Q}=0.02$. Grid size $500\times 500$, $t=1750000$.\label{f:lhdrop}}
\end{figure}

The two-dimensional spinodal decomposition was simulated taking into account the pressure work and
the latent heat of phase transition. The initial uniform fluid density was $\tilde{\rho}_0=1$, and
the temperature was everywhere $\tilde{T}_0=0.8$. Simulation results are shown in
Fig.~\ref{f:lh2d}. The temperature of vapor decresed to $\tilde{T}_v\approx 0.77$ which is lower
than the initial one, and the temperature of liquid reached the value $\tilde{T}_l\approx 0.83$
which is higher than the initial one due to the release of latent heat at the condensation.

In further evolution of the system, the number of droplets decreases due to the coalescence and the
evaporation of smaller droplets and growth of larger ones. The distribution of tempersture tends to
the uniform one due to heat conductivity. Figure~\ref{f:lhdrop} shows the stage when only two
droplets remain. The temperature of smaller droplet is lower due to the evaporation, the larger
droplet is heated due to the condensation. The temperature difference is $\Delta\tilde{T}\approx
0.01$. When the process ends, and only one droplet remains, the nonuniformity of the temperature
decreases to $\Delta\tilde{T}<0.001$.

\section{Conclusion}\label{concl}

The method of additional LB component was developed for multiphase thermal flows. The algorithm
takes into accont heat conduction, pressure work and latent heat of phase transition. The method is
interface-capturing, no tracking of phase boundaries and conditions at them is needed. Numerical
tests shows Galelean invariance of the method, low scheme diffusion of energy, isotropy and
stability. Calculated behavior of sound speed agrees well with theoretical predictions. The method
developed is applicable for simulating flows with heat and mass transport and phase transitions.

\section*{Acknowledgements}

This work was supported by the Russian Scientific Foundation, Grant N 16-19-10229.

%\bibliographystyle{unsrt}
%\bibliography{lattice,medvbibl}

\end{document}